\documentclass[		%comment out when using nature.cls
	aps,
	showpacs,
	superscriptaddress,
	preprintnumbers,
	amsmath,
	twoside,
	amsfonts,
	twocolumn,
	amssymb,
	prb]		%use 'pra' to get supplemental material right
{revtex4-1}
\usepackage[T1]{fontenc}
\usepackage[latin9]{inputenc}
\usepackage{graphicx}	%important to make '\includegraphics' operational with nature.cls

\usepackage{hyperref}
\usepackage{color}		%to use colours
\usepackage{amsmath}
\usepackage{amsfonts}
\usepackage{amssymb}
\usepackage{lmodern}		%allows using bold and italics at the same time

\def \d {^\dagger}

\def \action {\mathcal{S}}
\def \veck {\mathbf{k}}

\def \vecq {\mathbf{q}}

\def \sites {N}
\def \flavours {\mathcal{N}}
\newcommand{\eu}{^{}}	%eu stands for empty up
\newcommand{\ed}{_{}}	%ed stands for empty down

\newcommand{\eqn}[1] {eq.~(\ref{#1})}

\newcommand{\Fig}[1]{Fig.~\ref{#1}}

\newcommand{\SP}{P_s\eu}
\newcommand{\AP}{P_a\eu}
\newcommand{\SA}{A_s\eu}
\renewcommand{\AA}{A_a\eu}
\newcommand{\SPM}{\SP \text{-mode}}
\newcommand{\APM}{\AP \text{-mode}}
\newcommand{\SAM}{\SA \text{-mode}}
\newcommand{\AAM}{\AA \text{-mode}}

\newcommand{\MAPM}{M_{\AP}\eu}
\newcommand{\MSAM}{M_{\SA}\eu}
\newcommand{\MAAM}{M_{\AA}\eu}
\newcommand{\doping}{p}

\newcommand{\ourtitle}{Surprises in the $t$-$J$ model: Implications for cuprates}

\begin{document}

\title{\ourtitle}
%%%% comment out if using nature.cls %%%%%%%%%%%%%
\author{Aabhaas V. Mallik}
\email{aabhaas@iisc.ac.in}
\affiliation{Department of Physics, Center for Condensed Matter Theory, Indian Institute of Science, Bengaluru - 560012, India}
\author{Gaurav K. Gupta}
\email{ggaurav@iisc.ac.in}
\affiliation{Department of Physics, Center for Condensed Matter Theory, Indian Institute of Science, Bengaluru - 560012, India}
\author{Vijay B. Shenoy}
\email{shenoy@iisc.ac.in}
\affiliation{Department of Physics, Center for Condensed Matter Theory, Indian Institute of Science, Bengaluru - 560012, India}
\author{H. R. Krishnamurthy}
\email{hrkrish@iisc.ac.in}
\affiliation{Department of Physics, Center for Condensed Matter Theory, Indian Institute of Science, Bengaluru - 560012, India}
\date{\today}

%\pacs{74.20.-z, 74.72.-h, 74.90.+n}
%%%%%%%%%%%%%%%%%%%%%%%%%%%%%%%%%%%%%%%%%%%%%%%%%%

%%%% uncomment the below if using nature.cls %%%%%
%\author{Aabhaas V. Mallik$^{1}$, Gaurav K. Gupta$^{1}$, Vijay B. Shenoy$^{1}$ \& H. R. Krishnamurthy$^{1}$}
%\maketitle
%\begin{affiliations}
%\item Center for Condensed Matter Theory, Department of Physics, Indian
%Institute of Science, Bengaluru, India - 560012
%\end{affiliations}
%%%%%%%%%%%%%%%%%%%%%%%%%%%%%%%%%%%%%%%%%%%%%%%%%%
\begin{abstract}
	The $t$-$J$ model is a paradigmatic model for the study of strongly correlated electron systems. In particular, it has been argued that it is an appropriate model to describe the cuprate high-T$_c\eu$ superconductors. It turns out that a comprehensive understanding of the gamut of physics encoded by the $t$-$J$ model is still an open problem. In recent years some remarkable experiments on the cuprates, for example, discovery of nodeless superconductivity in underdoped samples (PNAS 109, 18332 (2012)), discovery of $s$-wave like gap in the pseudogap phase (Phys. Rev. Lett. 111, 107001 (2013)), and observation of polar Kerr effect (PKE) (Phys. Rev. Lett. 112, 047003 (2014)), have thrown up new challenges for this model. Here, we present results demonstrating that, within the slave-particle formulation of the t-J model, the $d$-wave superconductor is unstable at low doping to its own anti-symmetric phase mode fluctuations when the effect of fluctuations is treated self-consistently. We then show that this instability gives way to a time reversal symmetry broken $d + is$-SC in the underdoped region which has superfluid stiffness consistent with Uemura relation, even with a large pair amplitude. We show that our results are consistent with existing experiments on cuprates and suggest that Josephson (SQUID interferometry) experiments can clearly distinguish the $d+is$-SC from a host of other possibilities alluded to be contributing to the physics of underdoped cuprates. We also comment on other theoretical studies {\it vis-a-vis} ours.

\end{abstract}
\maketitle	%comment out if using nature.cls

The rich phase diagram of the copper oxide high-$T_c\eu$ superconductors has been a major source of inspiration for several fields of inquiry in both experimental and theoretical condensed matter physics\cite{KeimerNat15}. One of the important theoretical offshoots has been the extensive study of strongly correlated model systems like the $t$-$J$ model\cite{SpalekAPPA07,RicePRB88,AndersonJPCM04}. While this model has been quite successful in explaining several features of the cuprate phase diagram\cite{LeeRMP06,GrosAdvPhys07,FukuyamaRPP08}, a comprehensive understanding of the low temperature phase in the underdoped region is yet to be achieved\cite{SorellaPRB12,TroyerPRL14}. Recent experiments like the observation of nodeless superconductivity in the underdoped cuprates\cite{ShenPNAS12,RazzoliPRL13}, and, clear signatures of breaking of time reversal symmetry\cite{KapitulnikPRL14,FriedPRB14} raise new challenges. A natural question one would like to ask is how much of this new physics is within the scope of the $t$-$J$ model? The difficulty in addressing this question is two fold: 1) strong electron correlations, and, 2) increased importance of long-wavelength fluctuations because of low dimensionality. Numerical techniques like the Variational Monte Carlo (VMC) using projected mean field like wavefunctions\cite{OgataJPSJ96,PathakPRL09} or Projected Entangled Pair States (PEPS)\cite{TroyerPRL14} do a good job of accounting for strong correlations, but miss out on the long-wavelength fluctuations because of finite size limitations. The cluster DMFT\cite{KotliarPRL08} studies on the related Hubbard model have similar issues.

The slave-particle formulation of the $t$-$J$ model\cite{BaskaranPRB88,ReadJPC83} has an unique advantage in this respect. At the mean field level it agrees qualitatively with the VMC studies\cite{ParamekantiPRB04} indicating that the effect of strong correlations are getting accounted for. At the same time, being analytically tractable, it provides the framework for a systematic study of the effect of long-wavelength fluctuations beyond the mean field theory. It is worth noting that previous slave-particle studies of the $t$-$J$ model have mostly focused on incorporating the effect of fluctuations of the emergent gauge fields in the finite temperature phases, to better describe the effect of strong correlations\cite{LeePRB92,LeePRL09}. The description of the zero temperature phase in these studies are mean field like and, by and large, ignore the effect of fluctuations in the order parameter field. This is the crucial missing piece of the theory that we aim provide in the current work, leading to a set of intriguing conclusions.

Using the slave-particle formulation of the $t$-$J$ model, we first argue that at zero temperature the fluctuations in the pairing order parameter are likely to be the most relevant ones in the overdoped and the moderately underdoped regions of the phase diagram. We then, with model parameters relevant for cuprate superconductors, use a self consistent method\cite{DienerPRA08} to estimate the effect of these fluctuations on the saddle point order\footnote{From now on, unless otherwise mentioned, all our discussions will pertain to zero temperature quantities.}. We find that the results of this fluctuation consistent calculation are surprisingly different from their mean field counterparts. One of the key difference is that the $d$-wave superconducting order ($d$-SC) becomes unstable to its own fluctuations (unrelated to any non-pairing competing order) giving way to a $d+is$-SC for hole doping ($\doping$) $\lesssim 0.12$. While, a $d+is$-SC can naturally account for nodeless superconductivity, it also breaks microscopic time reversal ($\mathcal{T}$) symmetry\footnote{It also breaks $90^\circ\ed$ rotation symmetry, $\mathcal{R}$, while preserving the product of the two, $\mathcal{TR}$.}, a prerequisite for the observation of PKE\cite{FriedPRB14}. Other salient features of our fluctuation consistent theory which differ remarkably from the mean field theory are as follows: i) The value of hole doping on the overdoped side at which the $d$-SC subsides comes down noticeably to around $\doping \sim 0.33$, which is much closer to the experimentally observed values for cuprate superconductors. ii) On the underdoped side the $d+is$-SC order has a large extended $s$-wave amplitude, but the superfluid stiffness is approaching zero. This is consistent with the Uemura relation\cite{UemuraPRL89} and presents a new scope for understanding the observation of Nernst effect\cite{UchidaNature00,KingshukAnnPhys16} as arising from the physics of preformed pairs. iii) Going further in the underdoped side ($\doping \sim 0.055$), the fluctuation modes (except one) become soft and we find no uniform superconducting order in the $d$-wave, $s$-wave or $d+is$-wave channels. In the discussion section we give a more detailed comparison of our theory with some recent experiments, and, also propose experimental directions which can put our claims to test.

\section{Model}	%use \section*{} when using nature.cls
The $t$-$J$ model written in terms of the electron creation ($c_{i\sigma}\d$) and annihilation operators ($c_{i\sigma}\eu$) is as follows,
{\small
\begin{equation}
	\label{eq:H_tJ}
	H_{tJ}^{} = P \left[ - \sum_{\stackrel{i, \boldsymbol \delta}{\sigma}} t(\boldsymbol \delta) c_{i+\boldsymbol \delta \sigma} \d c_{i \sigma}^{} + J\sum_{\langle i,j \rangle} \left( \mathbf{S}_i \cdot \mathbf{S}_j - \frac{1}{4} n_i n_j \right) \right] P
\end{equation}}
where, $P$ is a projection operator which restricts $H_{tJ}^{}$ to the no-double-occupancy sector of the Hilbert space. $\mathbf{S}_i\eu$ and $n_i\eu$ are, respectively, the electron spin and number operators on the site $i$. $t(\boldsymbol \delta)$ is the hopping amplitude from a site $i$ to a neighbor at site $i+\boldsymbol \delta$. In our calculations we shall use the nearest neighbor hopping amplitude $t$, and, the next nearest neighbor hopping amplitude, $t' = -0.3t$. We shall also set the nearest neighbor exchange interaction energy, $J$, to $0.3t$.

The non-holonomic no-double-occupancy constraint ($n_i \le 1$) in the electron Hilbert space can be traded with a holonomic constraint for slave-particles defined as follows,
\begin{equation}
	c\d_{i\sigma} = f\d_{i\sigma} h_i\eu + \epsilon_{\sigma \sigma'}\eu f_{i\sigma'}\eu d_i\d
	\label{eq:slave_particles1}
\end{equation}
where $\epsilon_{\uparrow \downarrow}\eu = - \epsilon_{\downarrow \uparrow}\eu = 1$, and, $f_{i\sigma}\eu$, $h_i\eu$ and $d_i\eu$ are, respectively, the spinon (fermions with $z$-component of spin $\sigma$), holon (spinless bosons) and doublon (spinless bosons) annihilation operators at site $i$, such that
\begin{equation}
	\sum_\sigma f\d_{i\sigma} f_{i\sigma}\eu + h_i\d h_i\eu + d_i\d d_i\eu = 1
	\label{eq:slave_particles2}
\end{equation}
As the names suggest, the electronic states which have double occupancies correspond to some state in the slave-particle Hilbert space with non-zero number of doublons. Therefore, the no-double-occupancy constraint on the electron Hilbert space is equivalent to no-doublon constraint in the slave-particle Hilbert space, and, \eqn{eq:slave_particles1} and (\ref{eq:slave_particles2}) reduce to
\begin{equation}
	c\d_{i\sigma} = f\d_{i\sigma} h_i\eu
	\label{eq:slave_particles3}
\end{equation}
and
\begin{equation}
	\sum_\sigma f\d_{i\sigma} f_{i\sigma}\eu + h_i\d h_i\eu = 1
	\label{eq:slave_particles4}
\end{equation}
which is a convenient holonomic constraint that can be implemented using Lagrange multipliers, and hence, is analytically tractable. Of course, \eqn{eq:slave_particles3} is no longer an operator identity and for a consistent description of the $t$-$J$ model in terms of the spinons ($f$) and holons ($h$) one has to make sure that the operators involved have the same matrix elements in the two Hilbert spaces\cite{Kotliar2PRB88}. For example, the interaction term in \eqn{eq:H_tJ} is written as
{\small
\begin{multline}
	P \left( \mathbf{S}_i \cdot \mathbf{S}_j - \frac{1}{4} n_i n_j \right) P \rightarrow \\
	\frac{1}{4} \sum_{\stackrel{\sigma,\sigma',}{\tilde{\sigma},\tilde{\sigma}'}} \left( f_{i\sigma}\d \boldsymbol{\tau}_{\sigma \sigma'}\eu f_{i\sigma'}\eu \right) \left( f_{j\tilde{\sigma}}\d \boldsymbol{\tau}_{\tilde{\sigma} \tilde{\sigma}'}\eu f_{j\tilde{\sigma}'}\eu \right)
	- \frac{1}{4} \left( 1 - h_i\d h_i\eu \right)\left( 1 - h_j\d h_j\eu \right)
	\label{eq:interaction_term}
\end{multline}}
where $\boldsymbol{\tau}$ are the Pauli matrices.

The partition function in the grand canonical ensemble can, thus, be written in the path integral language as follows
\begin{equation}
	Z = \int \mathcal{D}f\d\ed\ \mathcal{D} f\ \mathcal{D} h^*\ed\ \mathcal{D} h\ \mathcal{D} \lambda\ \exp\left(-\int_{0}^\beta \text{d} \tau\ \mathcal{L}_{1}\eu \right)
	\label{eq:partition_func1}
\end{equation}
where $\beta$ is the inverse temperature, and,
\begin{equation}
	\mathcal{L}_{1}\eu = \mathcal{L}_t\eu + \mathcal{L}_h\eu + \mathcal{L}_f\eu 
	\label{eq:L_1}
\end{equation}
with
{\small
\begin{equation}
	\mathcal{L}_t\eu = -\sum_{\stackrel{i,\boldsymbol{\delta}}{\sigma}} t(\boldsymbol{\delta}) h_{i+\boldsymbol{\delta}}^* h_i\eu f_{i\sigma}\d f_{i+\boldsymbol{\delta}\sigma}
	\label{eq:L_t}
\end{equation}}
{\small
\begin{equation}
	\mathcal{L}_h\eu = \sum_i \left[ h_i^* \left( \frac{\partial}{\partial \tau} + \frac{3}{4}J - i\lambda_i\eu \right) h_i\eu - \frac{J}{4}\sum_{\hat{\alpha} \in \{ \hat{x}, \hat{y}\}} h_i^*h_i\eu h_{i+\hat{\alpha}}^*h_{i+\hat{\alpha}}\eu\right]
	\label{eq:L_b}
\end{equation}}
{\small
\begin{multline}
	\mathcal{L}_f\eu = \sum_i \left[ \sum_\sigma \eu f_{i\sigma}\d \left( \frac{\partial}{\partial \tau} - \mu_f\eu - i\lambda_i\eu \right) f_{i\sigma} \vphantom{\sum_{\stackrel{\hat{\alpha} \in \{ \hat{x}, \hat{y}\}}{\sigma,\sigma',\tilde{\sigma},\tilde{\sigma}'}}} \right. \\
	\left. + \frac{J}{4}\sum_{\stackrel{\hat{\alpha} \in \{ \hat{x}, \hat{y}\}}{\sigma,\sigma',\tilde{\sigma},\tilde{\sigma}'}} \left( f_{i\sigma}\d \boldsymbol{\tau}_{\sigma \sigma'}\eu f_{i\sigma'}\eu \right) \left( f_{i+\hat{\alpha}\tilde{\sigma}}\d \boldsymbol{\tau}_{\tilde{\sigma} \tilde{\sigma}'}\eu f_{i+\hat{\alpha}\tilde{\sigma}'}\eu \right) \eu\right]
	\label{eq:L_f}
\end{multline}}
, and, $\lambda_i$ being the Lagrange multiplier which implements the constraint \eqn{eq:slave_particles4}. $\mu_f\eu$ is the chemical potential which fixes the average number of spinons, and hence, the average number of holons in the system (cf \eqn{eq:slave_particles4}). One can immediately appreciate that the following transformations leave $\mathcal{L}_1\eu$ invariant
\begin{equation}
	f_{i\sigma}\eu \rightarrow \text{e}^{i \phi_i\eu}\ed f_{i\sigma}\eu \ \ , \ \ h_i\eu \rightarrow \text{e}^{i \phi_i\eu}\ed h_i\eu \ \ , \ \ \lambda_i\eu \rightarrow \lambda_i\eu + \frac{\partial \phi_i\eu}{\partial \tau}\ \ .
	\label{eq:gauge_transfm1}
\end{equation}
Clearly, $\mathcal{L}_1\eu$ has an emergent $U(1)$ gauge symmetry with the field $\lambda_i\eu$ transforming like the scalar potential. The phase of the fields corresponding to $\langle f_{i\sigma}\d f_{j\sigma}\eu \rangle$ and $\langle h_i^* \rangle \langle h_j\eu \rangle$ transform like the vector potential, $a_{ij}\eu$. To account for this redundancy, the path integral involved in the evaluation of the partition function in \eqn{eq:partition_func1} sums over each gauge invariant configuration of the involved fields only once.

At zero temperature, in the overdoped to moderately underdoped region, the holons are condensed and the Goldstone mode of this condensate along with the gauge fields (scalar and vector potentials) are gapped out by the Anderson-Higgs mechanism\cite{LeePRL09}. In this limit
\begin{equation}
	h_i\eu = (h_0\eu + \tilde{h}_i\eu) \ \exp\left(i\theta_i^{(h)}\right) \sim \sqrt{\doping}\ \exp\left(i\theta_i^{(h)}\right)
	\label{eq:holon_condensate1}
\end{equation}
where $\doping$ is the hole doping above half filling, $h_0\eu = \sqrt{\doping}$ is the amplitude of the holon condensate, $\tilde{h}_i\eu (\tau)$ is the corresponding amplitude fluctuation field, and, $\theta_i^{(h)}(\tau)$ is the phase field of the holon condensate. Further, it is convenient to use the gauge invariant fermionic field defined as follows,
\begin{equation}
	\tilde{f}_{i\sigma}\eu = f_{i\sigma}\eu \exp\left(-i\theta_i^{(h)}\right)\ \
	\label{eq:e-like1}
\end{equation}
which, like the electrons, would directly couple to the external electromagnetic field\footnote{The electron annihilation operator ($c_{i\sigma}\eu$) in the no-double occupancy sector is $c_{i\sigma}\eu \sim h_0\eu\tilde{f}_{i\sigma}\eu = \sqrt{\doping}\tilde{f}_{i\sigma}\eu$}. In terms of the fields $\tilde{f}$ and $\tilde{h}$, and, the Hubbard-Stratonovich fields $\Delta_{ij}\eu(\tau)$ and $K_{ij}\eu(\tau)$ introduced to decouple the quartic term in $\mathcal{L}_{\tilde{f}}\eu$ in the particle-particle and particle-hole channels, respectively, the partition function in \eqn{eq:partition_func1} becomes
{\small
\begin{equation}
	Z = \int \mathcal{D}\tilde{f}\d\ed\ \mathcal{D} \tilde{f}\ \mathcal{D} \Delta^*\ed \ \mathcal{D} \Delta\ \mathcal{D} \tilde{K}\ \mathcal{D} \tilde{a}\ \mathcal{D} \tilde{h}\ \mathcal{D} \tilde{\lambda}\ \exp\left(-\int_{0}^\beta \text{d} \tau\ \mathcal{L}_{2}\eu \right)
	\label{eq:partition_func2}
\end{equation}}
where $\tilde{\lambda}_i\eu (\tau) = \lambda_i\eu - \partial _\tau \eu \theta_i^{(h)} (\tau)$ and $\tilde{a}_{ij}\eu (\tau) = a_{ij}\eu (\tau) - (\theta_j^{(h)} (\tau) - \theta_i^{(h)} (\tau))$ are the gauge invariant `plasmon' fields, and, $\tilde{K}_{ij}\eu(\tau)$ is the amplitude of the particle-hole field $K_{ij}\eu (\tau) = \tilde{K}_{ij}\eu (\tau) \exp\left( i\tilde{a}_{ij}\eu (\tau) \right)$.
%\begin{equation}
%	\mathcal{L}_2\eu = \mathcal{L}_{\tilde{h},\tilde{\lambda},\tilde{a},\tilde{K}}\eu + \mathcal{L}_{\tilde{f},\tilde{h},\tilde{\lambda},\tilde{a},\tilde{K},\Delta}\eu \quad.
%	\label{eq:L_2}
%\end{equation}
$\mathcal{L}_{2}\eu$ describes the gapped amplitude excitations of the holon condensate and the Higgs'ed gauge fields (`plasmon' mode). It includes the amplitude excitations of the particle-hole field $K_{ij}(\tau)$ which are again expected to be gapped. But, most importantly, it also describes the fermionic sector along with the possibility of pairing and condensation.

In the presence of fermionic pairing the system has soft collective modes in the underdoped regime (see {\it e.g.} \cite{ShenoyEPL17}) which, given that all the other fields are gapped, might actually be the most important contributor to fluctuations about the mean field theory. With this in mind, we choose to work with the following construction: 1) We set the gapped fields $\tilde{h}$ and $\tilde{\lambda}$ to their mean field value of zero, and assume that the fluctuations in the pairing channel do not affect their status. 2) We, further, discount the fluctuations in the $K$ field as they are also expected to be gapped. 3) We analyze and incorporate the Gaussian fluctuations in the pairing channel $\Delta$ within a self-consistent scheme, described in the next section, to explore how do they effect the mean field pairing scenario.

\section{Fluctuation consistent saddle point}	%use \section*{} when using nature.cls
Within this construction the partition function in \eqn{eq:partition_func2} becomes
{\small
\begin{equation}
	Z = \int \mathcal{D}\tilde{f}\d\ed\ \mathcal{D} \tilde{f}\ \mathcal{D} \Delta^*\ed \ \mathcal{D} \Delta\ \mathcal{D} K^*\ed\ \mathcal{D} K\ \exp\left(-\int_{0}^\beta \text{d} \tau\ \mathcal{L}_{\tilde{f},K,\Delta}\eu \right)
	\label{eq:partition_func3}
\end{equation}}
where, we have replaced $\mathcal{D} \tilde{K}\ \mathcal{D} \tilde{a}$ in \eqn{eq:partition_func2} with $\mathcal{D} K^*\ed\ \mathcal{D} K$, and, in the momentum-Matsubara representation
\begin{widetext}
{\small
\begin{multline}
	\int _0^\beta \text{d} \tau\ \mathcal{L}_{\tilde{f},K,\Delta}\eu \equiv \action \left[\left\{ \tilde{f}_{k\sigma}\eu, \Delta_{q\alpha}\eu, K_{q\alpha}\eu \right\} \right] = \sum _{k,\sigma} \tilde{f}_{k \sigma}\d \left( -ik_n + \xi_\veck^{} \right) \tilde{f}_{k \sigma}^{} + \sum_{\stackrel{q}{\alpha \in \{\hat{x},\hat{y}\}}} \left[ \frac{|\Delta_{q\alpha}|^2\ed}{J_P\eu} - \frac{1}{\sqrt{\sites \beta}}\left( b_{q\alpha}\d \Delta_{q\alpha} + \Delta^*_{q\alpha} b_{q\alpha}\eu \right) \right] \\
	+ \sum_{\stackrel{q}{\alpha \in \{\hat{x},\hat{y}\}}} \left[ \frac{|K_{q\alpha}|^2\ed}{J_K\eu} - \frac{1}{\sqrt{\sites \beta}} \left( \chi_{q\alpha}\d K_{q\alpha}\eu + K^*_{q\alpha}\chi_{q\alpha}\eu \right) \right]
	\label{eq:action1}
\end{multline}}
\end{widetext}
with $k\equiv (\omega_n\eu=(2n+1)\pi \beta^{-1}\ed,\veck)$, $q\equiv (q_l\eu=2l\pi \beta^{-1}\ed,\vecq)$,
\begin{equation}
	\xi_\veck \eu = - 2 \doping t \left( \cos k_x\eu + \cos k_y\eu \right) - 4 \doping t' \cos k_x\eu \cos k_y\eu - \mu_f\eu
	\label{eq:dispersion}
\end{equation}
%\begin{eqnarray}
%	\nonumber
%	\Delta _{q\alpha}\eu & = & \frac{1}{\sqrt{\sites \beta}}\int _0^\beta \text{d} \tau\ \sum_i \text{e}^{i(q_l\eu\tau-\vecq\cdot\vecr_i\eu)}\ed \Delta_{i,i+\alpha}\eu (\tau)
%	\\
%	K _{q\alpha}\eu & = & \frac{1}{\sqrt{\sites \beta}}\int _0^\beta \text{d} \tau\ \sum_i \text{e}^{i(q_l\eu\tau-\vecq\cdot\vecr_i\eu)}\ed K_{i,i+\alpha}\eu (\tau)
%	\label{eq:HS_fields1}
%\end{eqnarray}
, and, $\sites$ being the number of lattice sites on the square lattice with periodic boundary condition.
%\begin{eqnarray}
%	\nonumber
%	b_{q\alpha}\d & = & \frac{1}{\sqrt{\sites \beta}}\int _0^\beta \text{d} \tau\ \sum_i \text{e}^{-i(q_l\eu\tau-\vecq\cdot\vecr_i\eu)}\ed \frac{1}{\sqrt{2}}\left( \tilde{f}_{i \uparrow}\d (\tau)\tilde{f}_{i+\alpha \downarrow}\d (\tau) - \tilde{f}_{i \downarrow}\d (\tau)\tilde{f}_{i+\alpha \uparrow}\d (\tau) \right)
%	\\
%	\chi_{q\alpha}\d & = & \frac{1}{\sqrt{\sites \beta}}\int _0^\beta \text{d} \tau\ \sum_i \text{e}^{-i(q_l\eu\tau-\vecq\cdot\vecr_i\eu)}\ed \frac{1}{\sqrt{2}}\left( \tilde{f}_{i \uparrow}\d (\tau)\tilde{f}_{i+\alpha \uparrow}\eu (\tau) + \tilde{f}_{i \downarrow}\d (\tau)\tilde{f}_{i+\alpha \downarrow}\eu (\tau) \right)\ .
%	\label{eq:pairing_field1}
%\end{eqnarray}
Note that, $J_P\eu=3J/4$ and $J_K\eu=3J/4$ are chosen in such a way that the saddle point theory of \eqn{eq:action1} agrees with an appropriate variational theory\cite{LeeRMP06}.

On integrating out the quadratic $\tilde{f}$ fields from \eqn{eq:partition_func3}, we obtain an effective action ($\tilde{\action}[\{\Delta_{q\alpha}\eu, K_{q\alpha}\eu \}]$) for the $\Delta$ and $K$ fields. The saddle point equations obtained by setting $\partial \tilde{\action} / \partial \Delta^*_{q\alpha} = 0$ and $\partial \tilde{\action} / \partial K^*_{q\alpha} = 0$, and, using the following ansatz for the saddle point configuration of the fields
{\small
\begin{equation}
	\begin{matrix}
	\left( \Delta_{qx}\eu \right)_{SP}\eu & = & \sqrt{\sites \beta}(\Delta_d\eu + i \Delta_s\eu) \delta_{q,0} & \equiv & \sqrt{\sites \beta}\Delta_{SP}\eu \delta_{q,0} \\
	& & & & \\
	\left( \Delta_{qy}\eu \right)_{SP}\eu & = & - \sqrt{\sites \beta}(\Delta_d\eu - i \Delta_s\eu) \delta_{q,0} & = & - \sqrt{\sites \beta}\Delta_{SP}^{*} \delta_{q,0} \\
	& & & & \\
	\left( K_{q\alpha}\eu \right)_{SP}\eu & = & \sqrt{\sites \beta} K_{SP}\eu \delta_{q,0} & & \quad \quad \quad \quad \forall \ \alpha
	\end{matrix}
	\label{eq:SP_ans}
\end{equation}}
are
{\small
\begin{equation}
	\Delta_d\eu = \frac{J_P\eu}{2\sqrt{2}\sites}\sum_\veck \left( \cos k_x\eu - \cos k_y\eu \right) \frac{\Delta_{d,\veck}\eu}{E_\veck\eu}\tanh\left( \frac{\beta E_\veck\eu}{2} \right)
	\label{eq:Ddeq}
\end{equation}
\begin{equation}
	\Delta_s\eu = \frac{J_P\eu}{2\sqrt{2}\sites}\sum_\veck \left( \cos k_x\eu + \cos k_y\eu \right) \frac{\Delta_{s,\veck}\eu}{E_\veck\eu}\tanh\left( \frac{\beta E_\veck\eu}{2} \right)
	\label{eq:Dseq}
\end{equation}
\begin{equation}
	K_{SP}\eu = - \frac{J_K\eu}{2\sqrt{2}\sites}\sum_\veck \left( \cos k_x\eu + \cos k_y\eu \right) \frac{\tilde{\xi}_\veck\eu}{E_\veck\eu}\tanh\left( \frac{\beta E_\veck\eu}{2} \right)
	\label{eq:KSPeq}
\end{equation}}
where,
\begin{equation}
	\label{eq:tilde_xi_k}
	\tilde{\xi}_\veck\eu = \xi_\veck\eu - K_\veck\eu
\end{equation}
\begin{equation}
	\label{eq:K_k}
	K_\veck\eu = \sqrt{2}K_{SP}\eu (\cos k_x\eu +\cos k_y\eu)
\end{equation}
\begin{multline}
	\label{eq:Delta_k}
	\Delta_\veck\eu = \sqrt{2}\Delta_{d}\eu (\cos k_x\eu -\cos k_y\eu) + i\sqrt{2}\Delta_{s}\eu (\cos k_x\eu +\cos k_y\eu) \\
	\equiv \Delta_{d,\veck}\eu + i\Delta_{s,\veck}\eu .
\end{multline}
and
\begin{equation}
	\label{eq:E_k}
	E_\veck \eu= \sqrt{\tilde{\xi}_\veck^{2} + |\Delta_{\veck}|^{2}\ed}
\end{equation}
The saddle point equations have to be solved self-consistently along with the number equation
\begin{equation}
	\label{eq:mu_eq}
	\sites(1-\doping) = -\frac{\partial F}{\partial \mu}
\end{equation}
with $F$ being the grand canonical free energy. Typically, one replaces $F$ in \eqn{eq:mu_eq} by its saddle point estimate ($F_{SP}\eu$) ignoring the contribution from the fluctuations about the saddle point. A more accurate, and yet, tractable way is to include the effect of Gaussian fluctuations of the Hubbard-Stratonovich fields\cite{DienerPRA08}. In what follows we give the analytic expressions for $F_{SP}\eu$ and $F_{GF}\eu$, where $F_{GF}\eu$ is the grand free energy with contribution from Gaussian fluctuations of the pairing fields ($\Delta_{q\alpha}\eu$) included along with the saddle point contribution ($F_{SP}\eu$).

To obtain the grand free energy with the contribution of Gaussian fluctuations in the pairing field $\Delta_{q\alpha}\eu$, we need to start with the Gaussian action of the fluctuation fields $\eta_{q\alpha}\eu \equiv \Delta_{q\alpha}\eu - \left( \Delta_{q \alpha}\eu \right) _{SP}\eu$
\begin{equation}
	\label{eq:S_Gauss1}
	\tilde{\action}_G\eu[\left\{\eta_{q\alpha}^{*},\eta_{q\alpha}\eu \right\}] = \tilde{\action}_{SP}\eu - \sum_{q} \Lambda _q\d \mathbb{D} (q)^{-1}\ed \Lambda _q\eu
\end{equation}
where
{\small
\begin{multline}
	\frac{\tilde{\action}_{SP}\eu}{\beta \sites} = \frac{F_{SP}\eu}{\sites} = \frac{2|\Delta_{SP}\eu|^2\ed}{J_P\eu} + \frac{2|K_{SP}\eu|^2\ed}{J_K\eu} + \frac{1}{\sites}\sum_\veck \left( \tilde{\xi}_\veck\eu - E_\veck\eu \right) \\
	- \frac{2}{\sites \beta} \sum_\veck \ln \left( 1 + \text{e}^{-\beta E_\veck\eu}\ed \right)
	\label{eq:F_SP}
\end{multline}}
\begin{equation}
	\label{eq:fluc_vector}
	\Lambda \d _q = \left( \eta _{q0} ^{*} \quad \eta _{-q0} \eu \quad \eta _{q1} ^{*} \quad \eta _{-q1} \eu \right)
\end{equation}
and $\mathbb{D} (q)^{-1}\ed$ is the inverse fluctuation propagator. Then, formally,
\begin{equation}
	\label{eq:F_GF}
	\frac{F_{GF}\eu}{\sites} = \frac{F_{SP}\eu}{\sites} + \frac{1}{2\beta \sites} \sum_{q} \ln \left( \det \mathbb{D} (q)^{-1}\ed \right)
\end{equation}
\emph{Numerical complexity:} We would like to point out that the computation of the second term in the RHS of \eqn{eq:F_GF} is technically challenging. It involves integration of functions with integrable singularities arising from the nodal nature of the $d$-SC. We use an adaptive integration algorithm which is well suited to evaluate such integrals and provides us with systematically convergent results. Apart from these, some parts of this term appear to be divergent. But, on careful tracking of convergence factors which arise naturally in the path integral formulation, these can be handled analytically to give convergent physical results. %(details in Supplemental Material (SM)).

\section{Results}	%use \section*{} when using nature.cls
\begin{figure}
	\includegraphics[width=1.0\linewidth]{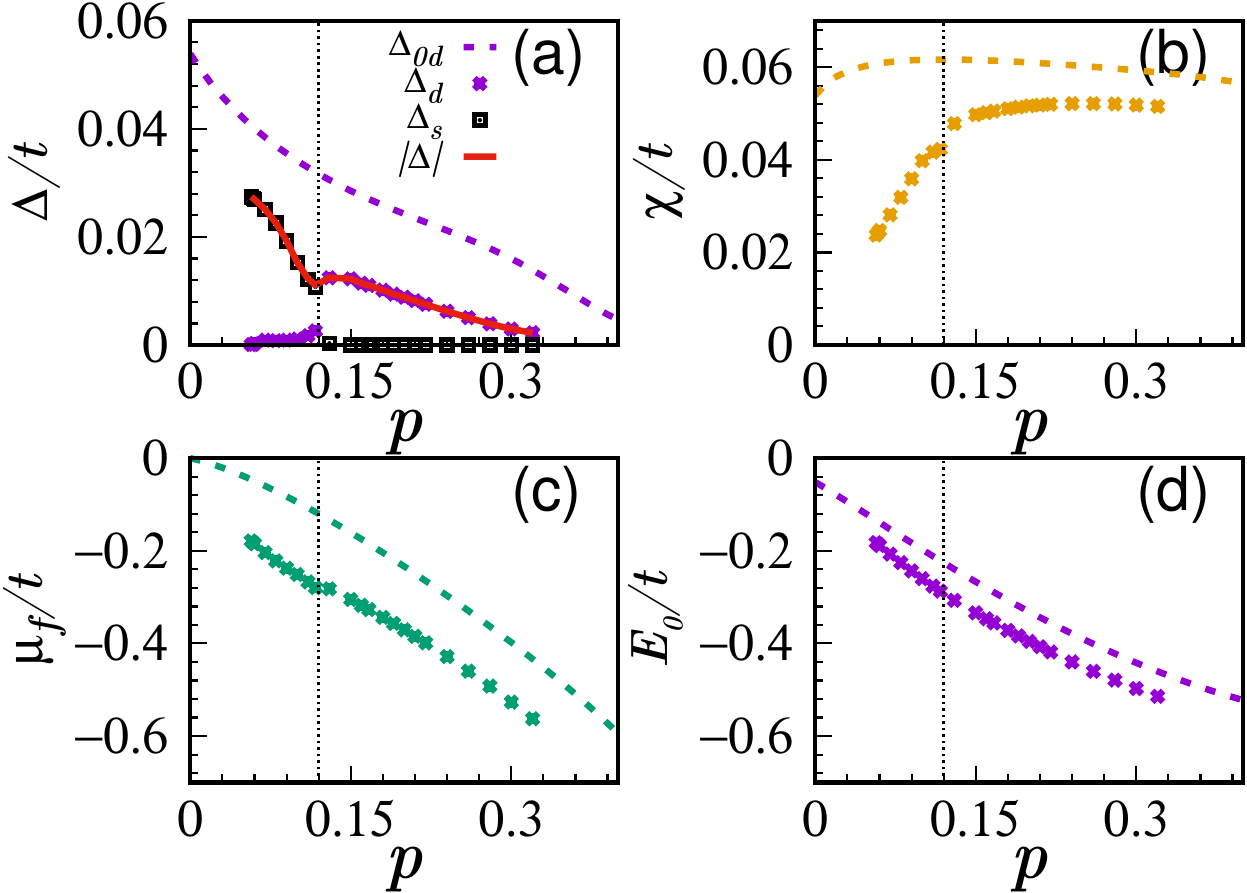}
	\caption{(color online) {\bf Fluctuation consistent saddle point:} (a) shows the pairing amplitude ($|\Delta| = \sqrt{\Delta_d^2 + \Delta_s^2}$) and its extended $s$-wave and $d$-wave components. (b) has the extended-$s$ bond kinetic order $K_{SP}$. (c) shows the variation of the chemical potential $\mu_{f}\eu$. And, (d) has the energy of the state corresponding to the saddle point. In all the plots above, the corresponding quantities in the usual saddle point theory are depicted for comparison by dashed curves. The dotted vertical line marks the doping value ($\sim 0.12$) where the fluctuation consistent theory has a phase transition.}
	\label{fig:figure1}
\end{figure}

The results obtained by solving the saddle point equations \eqn{eq:Ddeq}-(\ref{eq:KSPeq}) along with the number equation \eqn{eq:mu_eq} and \eqn{eq:F_GF} self-consistently are plotted in \Fig{fig:figure1} as a function of hole doping $\doping$. We represent all the quantities in units of the nearest neighbor hopping amplitude $t$. The dashed line in the plots show the usual saddle point results, {\it i.e.}, when $F$ in \eqn{eq:mu_eq} is replaced by $F_{SP}\eu$, while the points show the results at the fluctuation consistent saddle point. In \Fig{fig:figure1}(a) we have plotted the pairing amplitude and its different components with and without the inclusion of the effect of fluctuations. $\Delta_{0d}\eu$ is the $d$-wave pairing amplitude when the effect of fluctuations are not included. This is the usual slave-particle mean field result. $\Delta_d\eu$ and $\Delta_s\eu$, indicated respectively by purple crosses and black squares, are the $d$-wave and extended $s$-wave components of the $d+is$ pairing amplitude when the effect of fluctuations are included self-consistently. Interestingly, we find that there is a phase transition from a $d$-wave superconductor on the overdoped side to a $d+is$ superconductor on the underdoped side (with predominantly extended $s$-wave character), at $\doping_c\eu \sim 0.12$ marked by a vertical dotted line. Further, $|\Delta|= \sqrt{\Delta_d^2+\Delta_s^2}$, the magnitude of the pairing amplitude with effect of fluctuations included, is shown by a solid red line.

In \Fig{fig:figure1}(b) we have plotted the bond kinetic order $K$ as a function of $\doping$. The bond kinetic order is assumed to have an extended-$s$ symmetry (cf \eqn{eq:SP_ans} and (\ref{eq:KSPeq})). Therefore, it is not surprising that as the extended-$s$ pairing order ($\Delta_s\eu$) develops for $\doping < \doping_c\eu$, the bond kinetic order exhibits a fall. This is so because both these orders have weight in the same region of the Brillouin zone, and hence, compete to engage a common set of fermions in forming their respective orders. In \Fig{fig:figure1}(c) we have plotted the chemical potential $\mu_f\eu$ against $\doping$. While, in \Fig{fig:figure1}(d) the energy ($E_0\eu$) of the state of the system at the saddle point has been plotted. It is reassuring to note that the inclusion of the fluctuation contribution lowers the value of $E_0\eu$ compared to when they are not included. Further, there is no detectable kink in $E_0\eu(\doping)$, indicating that the transition at $\doping_c\eu$ is either a continuous phase transition or is weakly first order. Again, The dotted vertical line in all these plots marks the doping ($\sim 0.12$) where the transition in the pairing channel occurs.

To understand this transition better we analyze the slow and long-wavelength collective modes of the pair condensate. This is done by analyzing the inverse fluctuation propagator $\mathbb{D}^{-1}\ed (q)$ in the $q\rightarrow 0$ limit. Generically, this system has two phase modes and two amplitude modes corresponding to the complex valued fluctuations on the unique $x$ and $y$ bonds associated to each lattice site. On the overdoped side, where there is only $d$-wave pairing, the normal modes at $q=0$ are given by the symmetric and anti-symmetric combinations of the phase (amplitude) modes:  the symmetric phase mode ($\SPM$, which is gapless in the absence of coupling to the electromagnetic gauge field), the anti-symmetric phase mode ($\APM$), the symmetric amplitude mode ($\SAM$), and, the anti-symmetric amplitude mode ($\AAM$)\cite{ShenoyEPL17}. In the underdoped region where $d+is$ pairs stabilize the $\SPM$ and $\AAM$ continue to be the normal modes in the $q\rightarrow 0$ limit, while the $\APM$ and $\SAM$ combine to give two new normal modes. We continue to use the old nomenclature for these new modes; the mode whose gap parameter\footnote{The action of a bosonic mode $\phi$, in the slow and long-wavelength limit has the following form, $\phi^{*}\ed (q) \left[ a - b(iq_l\eu )^2\ed + c|\mathbf{q}|^2\ed \right] \phi(q)$. Here, $\sqrt{a/b}$ is the gap in the excitation of the mode. We call $a$ as the gap parameter and find it sufficiently useful for our discussions.} ($M$) connects with the $\APM$ of the $d$-wave condensate at the transition point is still called the $\APM$, and, the $\SAM$ of the $d+is$ condensate is defined in an analogous way.

\begin{figure}
	\includegraphics[width=1.0\linewidth]{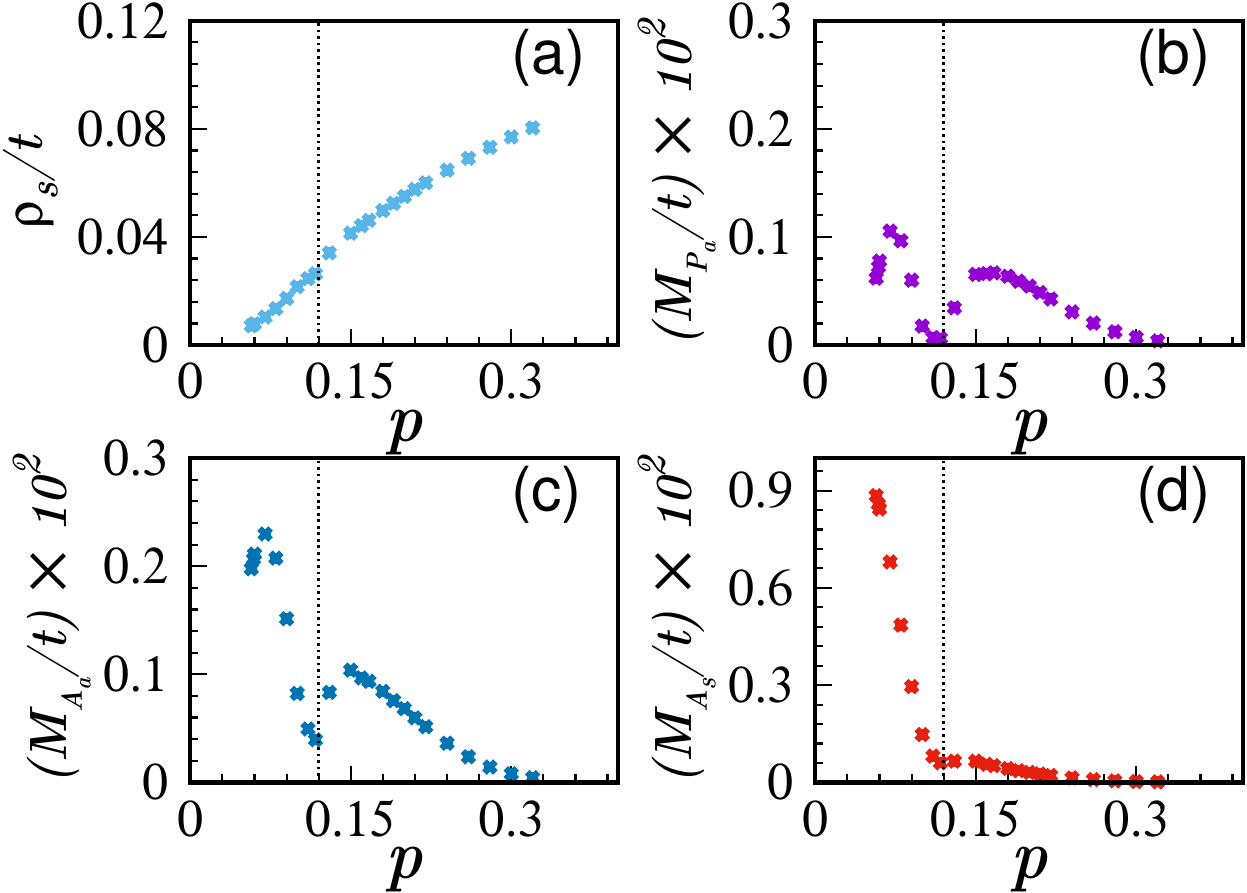}
	\caption{(color online) {\bf Properties of the collective modes:} All the data presented are obtained at the fluctuation consistent saddle point and the vertical dotted line marks the doping ($\sim 0.12$) where the fluctuation consistent theory hosts a phase transition. (a) shows the stiffness of the $\SPM$ mode. (b) has the gap parameter of the $\APM$. (c) depicts the gap parameter of the $\AAM$. And, (d) has the gap parameter of the $\SAM$.}
	\label{fig:figure2}
\end{figure}

In \Fig{fig:figure2} we plot the relevant properties of the collective modes of the pair condensate as a function of $\doping$, when the effect of the pairing field fluctuations are treated self-consistently. The vertical dotted line in all these plots is at $\doping_c\eu \sim 0.12$ where \Fig{fig:figure1}(a) hosts a phase transition. \Fig{fig:figure2}(a) shows the stiffness ($\rho_s\eu$) of the $\SPM$ which is the gapless Goldstone mode. It is remarkable that $\rho_s\eu$ goes to zero at $\doping \sim 0.055$ and concurs to the spirit of Uemura relation observed in experiments. In \Fig{fig:figure2}(b) we have the gap parameter, $\MAPM$, of the $\APM$. In \Fig{fig:figure2}(c) the gap parameter of the $\AAM$, $\MAAM$, has been plotted. And, the gap parameter, $\MSAM$, of the $\SAM$ is in \Fig{fig:figure2}(d). The phase transition from being a $d$-wave superconductor to a $d+is$ superconductor is marked by an instability of the $\APM$, as is clearly indicated by its vanishing gap parameter in \Fig{fig:figure2}(b). This also provides a good {\it a posteriori} justification of our choice to include only the fluctuations in the pairing channel in our fluctuation consistent theory. Further, we find that below $\doping \sim 0.055$ there is no stable uniform superconducting phase in any of the pairing channels, extended-$s$, $d$ and $d+is$, that we explore. This is marked by vanishing stiffness of the $\SPM$ and by the softening of the $\APM$ and the $\AAM$. Our theory does not have a good description of the state of the system below $\doping \sim 0.055$, and hence, we are unable to comment on the nature of the transition at this doping.

\section{Discussion}	%use \section*{} when using nature.cls
Let us begin with a discussion of our results on the overdoped side {\it vis-a-vis} other theories of the $t$-$J$ model and the large-$U$ Hubbard model, along with the experimental observations in cuprate superconductors. In the overdoped region the stiffness of the holon condensate is expected to be large and, therefore, our fluctuation consistent theory would also be expected to be most dependable in this region. Here, the $d$-wave pairing gap is known to grow with underdoping in both VMC\cite{ParamekantiPRB04} (compare with the variational parameter $\Delta_{var}\eu$) and slave-particle mean field theory\cite{LeeRMP06}. This is in contrast with our results where the $d$-wave pairing gap\footnote{The pairing scales plotted in \Fig{fig:figure1}(a) are directly related to the momentum dependent gap in the spectral function of the electron, but the electron pairing order parameter is obtained only after multiplying these by $\doping/J_P\eu$.} first increases and then attains a plateau at $\doping \sim 0.15$. Interestingly, this behavior of the $d$-wave pairing gap on the overdoped side agrees qualitatively with cluster DMFT studies of a closely related Hubbard model\cite{KotliarPRL08} and a large-$\flavours$ theory that we had presented in a similar setting\cite{ShenoyEPL17}. Most importantly, ARPES experiments on the cuprate Bi2212\cite{ShenPNAS12} and STM experiments on several cuprates\cite{SawatzkyRPP08} do find results consistent with such a behavior of the superconducting gap on the overdoped side. All these results put together, clearly suggest that the pseudogap is distinct from the gap arising because of $d$-wave pairing\cite{HoffmanScience14}. Yet another quantitative improvement that our fluctuation consistent theory provides over the slave-particle mean field theory is to pull down the value of doping at which the $d$-SC dies out (see \Fig{fig:figure1}(a)). This, again, compares favorably with the observed behavior in overdoped cuprate superconductors.

In the moderately underdoped region our results are remarkably different from other theories of the $t$-$J$ model. We find that on approaching the underdoped regime from the overdoped side, the $d$-SC makes a transition to a $d+is$-SC at $\doping_c\eu \sim 0.12$. This difference from other numerically exact theories like cluster DMFT and variational calculation using PEPS is, most likely, arising because of their severe finite size limitations, and hence, their inability to handle long-wavelength fluctuations which are crucial to stabilize the $d+is$-SC. Our fluctuation consistent theory does not have this limitation. We also point out that we evaluate all the momentum sums as Brillouin Zone integrals, and hence, capture the contribution of all long-wavelength Gaussian fluctuations correctly in the infinite system size limit. Here, we would like to note that a variational study of the $t$-$J$-$J'$ model, where $J' = J/2$ is the next nearest neighbor anti-ferromagnetic coupling, using fermionic PEPS finds that a $d+is$-SC is indeed a good candidate for the ground state wavefunction for all values of doping\cite{PoilblancPRB14}.

The $d+is$-SC, generically, does not have a node at the Fermi surface. It is very encouraging that such a phase is indeed observed in ARPES experiments on underdoped cuprate superconductors\cite{ShenPNAS12,RazzoliPRL13}. It is to be noted that the exact value of doping at which the $d$-SC to $d+is$-SC transition happens in our theory, will depend upon the details of the Fermi surface (choices of $t'$ and next to next nearest neighbor hopping $t''$). In the literature, however, the observation of nodeless superconductivity in underdoped cuprates has been attributed to a variety of other possibilities, like, superconductivity with pair wavefunction having $d+id$ symmetry\cite{RazzoliPRL13}, a topological $p+ip$ superconductor\cite{DHLeeNatPhys14,TanmoyarXiv14}, the presence of incommensurate spin density waves\cite{RazzoliNatCommun14,AtkinsonPRL12}, or just the presence of Coulomb disorder\cite{ChenPRB09}. Here, it is interesting to note that the phase sensitive experiments like the Josephson junction interferometry experiments\cite{WollmanPRL93,WollmanPRL95} or the ring magnetometry experiments\cite{TsueiPRL94} which were instrumental in confirming that cuprates are $d$-wave superconductors have mostly been applied to moderately underdoped to overdoped cuprates\cite{TsueiPRL04}. We propose that such phase sensitive experiments performed on the deeply underdoped regime can clarify much of the confusion regarding the symmetry of the pairing order parameter. While such experiments would clearly distinguish the different pairing symmetry possibilities, more careful study will be required to understand how spin density waves or Coulomb disorder may modify the results expected from a pure $d$-wave paired system. We would also like to point out that in the presence of disorder intrinsic to cuprates, a $d+is$ superconductor would generate spontaneous magnetization (which may contribute to the spin glass physics observed in some cuprates\cite{NiedermayerPRL98}). By producing controlled defects, this fact can also be used to distinguish a $d+is$ superconductor from a $d$-wave superconductor\cite{ChubukovPRB16}.% Further, the $d+id$ or $p+ip$ superconductors are topological and hence will have robust gapless edge states which, again, can be looked for in ARPES or STM experiments.

We have also computed the transverse conductivity, $\sigma_{xy}\eu (\veck = 0, \omega)$, to find that its imaginary component is zero for all $\omega$ and for all dopings of our interest. This implies that the $d+is$ state, which does not have time reversal symmetry, is not sufficient to give rise to the polar Kerr effect signal seen in experiments\cite{KapitulnikPRL14}. This is so because the uniform $d+is$ state preserves $\mathcal{T} \mathcal{R}$, where $\mathcal{T}$ is the time reversal operator and $\mathcal{R}$ rotates the system anti-clockwise by $90^\circ\ed$. This brings forth two interesting possibilities. 1) A charge density wave order which develops close to $\doping \sim 0.12$ breaks the point group symmetry of the $d+is$ state, and hence, gives rise to the observed polar Kerr effect. 2) A modulated $d+is$ superconducting saddle point (cf pair density wave), rather than one with the uniform $d+is$ pairing order, has lower energy. Both these possibilities are consistent with the observation that the polar Kerr effect onset temperature coincides with the onset of charge density wave order\cite{KapitulnikPRL14}. It is a matter of further investigation that which of the two scenarios is actually realized in cuprate superconductors and which of them can be rationalized within the $t$-$J$ model\cite{HirschfeldNJP17,LeearXiv18}.

With further underdoping, our fluctuation consistent theory predicts that the extended $s$ component of the pairing gap increases to significantly higher values, while the $d$ component diminishes completely by $\doping \sim 0.055$. This would imply that if the pseudogap phase in underdoped cuprate superconductors has contribution from preformed pairs, those pairs are ought to be of the extended $s$ type rather than of $d$ type. Further investigations will be required to see whether this scenario can help explain the observation of Nernst effect in pseudogap phase of the cuprates\cite{UchidaNature00,KingshukAnnPhys16}. Another outstanding feature of our theory is that the superfluid stiffness $\rho_s$ vanishes at around the same value of hole doping ($\sim 0.055$), which is consistent with the Uemura relation observed in experiments on several underdoped cuprates\cite{UemuraPRL89}. Further, the mass parameters corresponding to the $\APM$ and the $\AAM$ also approach zero around the same doping ($\sim 0.055$). Consequently, uniform superconductivity does not survive in any of the pairing channels investigated, namely, $d$, extended-$s$ or $d+is$, below this doping. This, again, is consistent with experiments on cuprates and leaves scope for other non-superconducting phases observed close to half filling. However, a note of caution must be added here. With underdoping fluctuations of the internal gauge fields will become more and more important and may lead to qualitative modifications of our theory in the deeply underdoped regime. Addressing this issue is outside the scope of the theory presented in this paper, and is a promising future work.

Overall, our fluctuation consistent theory brings in several insights beyond mean field slave-particle theory and numerical methods like the VMC and cluster DMFT. We hope this will prompt further investigations along these lines for other interacting two dimensional systems of interest, for example, the iron pnictide superconductors. Finally, we would like to end with a set of questions which our study throws up and whose answers can be very insightful in the context of cuprates: What is the nature of the state at finite temperatures on the underdoped side in the presence of the $d+is$ pairing with predominantly extended-$s$ character? What, if any, features of the cuprate pseudogap may be understood within this scenario\cite{FujimoriarXiv18}? What is the nature of the transition at $\doping_c\eu \sim 0.12$? If it is a continuous phase transition, what are the features of the associated criticality? What does this physics mean for the strange metal phase? Whether the non-pairing competing/intertwined orders scenario is really central in understanding the underdoped cuprates, or, are they just arising opportunistically close to the critical doping ($\doping_c\eu \sim 0.12$) where the pairing channel undergoes a transition? What happens in deeply underdoped regime $\doping \sim 0.055$? We hope to address some of these questions in our future work.

	{\bf Acknowledgements:} AVM thanks Subhro Bhattacharjee and Sumilan Banerjee for useful comments and suggestions. AVM also thanks Philippe Corboz for useful discussions and encouragement. We also acknowledge the following funding agencies of Govt. of India: SPMF/CSIR and DST.	%comment out if using nature.cls

\bibliography{bib_dpis}

%%%%%%%% uncomment if using nature.cls %%%%%%%
%\begin{addendum}
%\item We thank Subhro Bhattacharjee and Sumilan Banerjee for useful comments and suggestions. AVM would like to thank Philippe Corboz for useful discussions and encouragement. We also acknowledge the following funding agencies of Govt. of India: SPMF/CSIR and DST.
%\item [Competing Interests] The authors declare that they have no competing financial interests. 
%\item [Correspondence] Correspondence and requests for materials should be addressed to A. V. Mallik~(email: aabhaas.iiser@gmail.com).
%\end{addendum}
%%%%%%%%%%%%%%%%%%%%%%%%%%%%%%%%%%%%%%%%%%%%%%

\end{document}